DFTUZ95/18

# Hot and repulsive traffic flow

I. Campos and A. Tarancón,
*Departamento de Física Teórica, Universidad de Zaragoza,*
*50009 Zaragoza, Spain,*
*(e-mail isabel, tarancon, @sol.unizar.es)*

F. Clérot
*France Télécom, Centre National d'Etudes des Télécommunications*
*2 avenue Pierre Marzin, 22307 Lannion Cedex, France,*
*(e-mail fabrice.clerot@lannion.cnet.fr)*

L. A. Fernández
*Departamento de Física Teórica, Universidad Complutense de Madrid,*
*28040 Madrid, Spain,*
*(e-mail: laf@lattice.fis.ucm.es)*

**Abstract**

We study a message passing model, applicable also to traffic problems. The model is implemented in a discrete lattice, where particles move towards their destination, with fluctuations around the minimal distance path. A repulsive interaction between particles is introduced in order to avoid the appearance of traffic jam. We have studied the parameter space finding regions of fluid traffic, and saturated ones, being separated by abrupt changes. The improvement of the system performance is also explored, by the introduction of a non-constant potential acting on the particles. Finally, we deal with the behavior of the system when temporary failures in the transmission occurs.

# 1 Introduction

The formulation of adequate models for describing message passing through a Computer Network [1], the traffic flow in a city [2], adsorption of molecules in a crystal [3], or the mechanism of High Temperature Superconductors [4], is getting increasing attention in the last years.

All these problems share that they can be studied from the point of view of the Random Walk Theory. Nevertheless, the high number of degrees of freedom makes infeasible to extract relevant information from analytical calculations. The alternative is the use of numerical simulations.

One can think in formulating models, as close to the reality as possible, including as many degrees of freedom as one is able to handle. The resulting approaches are too complicate to study, not only in what concerns the description of relevant phenomena, but also because they lack of predictive power.

An alternative approach is to go to simpler formulations, less realistic, but retaining the main features of the physical system. Such models are described by few parameters, being in this way easier to handle, and a global study of the parameter space is feasible, without losing physical intuition.

Our hope is to obtain relevant results on complex systems by studying relatively simple models as is done in Statistical Mechanics (SM) where systems as complex as a real ferromagnetic lattice share many properties with models as simple as the Ising Model. In SM the relation between both systems is understood by means of such techniques as the Renormalization Group.

Unfortunately we are far from proving that kind of relation for traffic models but the study of simple model is anyway necessary.

When we deal with problems involving information flow, it is necessary to face with problems related with congestion processes. These systems undergo a transition from a situation of fluid traffic to another one characterized by the traffic jam. This change, can be related to a phase transition in SM, and the tools developed for statistical systems used. The first of all is to search for the variables governing this transition.

The transmission in the lattice, is conditioned by the maximum number of particles that a site is able to manage, the interaction between particles, and the geometric constraints such as the coordination number (number of neighbors). In what concerns the interaction, as said before, we hope that only the general features will be relevant.

In our model we implement movement by assigning to each particle a final point. In real traffic, the particles are often stopped by obstacles in their way. These obstacles are simulated by limiting the number of particles that a site can buffer simultaneously.

At this point, we return to the discussion of the problem from the point of view of SM. Let us focus on the situation where the density is high. It is intuitively clear that a particle should avoid zones of traffic jam by allowing it to surround obstacles. The system will be globally clearer, and we expect the throughput to be improved, in spite of the temporary misrouting of the particles. This idea is not only beneficial in situations of traffic jam, but also in general: allowing fluctuations around the minimal distance path improves the characteristics of the traffic.

In terms of SM, the above discussion means that by heating the system, ($T \neq 0$), we will obtain better performances [5].

In what concerns the interaction between particles, up to now we have introduced a contact interaction, the existence of obstacles, as sites completely occupied. The consequences of the contact interaction can be seen as an infinite potential acting on saturated sites.

An appropriate interaction term, is obtained by assigning an *electrical* charge to every particle. The repulsion will move the particles away from regions of high charge.

Repulsive forces are usually considered in the case of directed polymers [6].

We have introduced a term to control the size of the thermal fluctuations, and another one to simulate a repulsive force. The parameters controlling these terms are the temperature and the charge respectively. We will see how, by tuning both parameters, is possible to improve the performance of the transmission.

# 2 The Model

We consider a two-dimensional lattice with coordination number 4, and periodic boundary conditions. The particles live in the lattice sites, labeled by $n \equiv (n_0, n_1)$ and can move from a site to one of its 4 nearer neighbors in every time step. The maximum number of particles that a site can buffer will be denoted by $B$, and will be kept fixed along all the simulation at $B = 5$. We denote the occupation number of the site $n$ by $\sigma(n)$. In this notation, the particles are prevented of moving to sites with $\sigma(n) = B$. In addition we consider an infinite queue at each site for the messages waiting to be put on the network.

By analogy with SM systems, we will work with the inverse of the temperature, $\beta$. We denote the charge of the particles by $\kappa$ and the probability of particle injection to the lattice by $p$.

The dynamic of the system is the following:

1. We choose at random a lattice site.

2. A particle is added, with fixed probability $p$, to the queue of the site, waiting for being introduced in the lattice.

3. If the queue of the site is not empty, and $\sigma(n) < B$, a new particle is introduced in the lattice, and an endpoint is assigned to it randomly.

4. All particles in the considered site, try to move towards one of its 4 neighbors. For a given particle sitting on $n$ we must assign a probability of jumping to each of its neighbors. This probability is given by:

$$P(\pm\mu) = N \exp(\pm\beta \text{sign}({n^f}_\mu - n_\mu) - \kappa\sigma(n_\mu)) \ , \tag{1}$$

   where $n_\mu$ means the $\mu$ coordinate of the site $n$, and $n^f$ is the endpoint of the particle considered, and $N$ is the normalization constant. To choose between possible destinations we use a *Heat Bath* [7] algorithm.

5. The movement is granted if the chosen site has $\sigma(n) < B$. Otherwise the particle remains at its original site until a new movement attempt in the next iteration.

6. The particle disappears when moving to its endpoint.

In this way, the lattice sites simulate nodes of transmission and reception of messages coming up from the outside.

The factor multiplying $\beta$, is a potential term. It implies a constant force acting on the particle and driving it to its endpoint. The $\kappa$ term produces repulsion between particles sitting in nearer neighbors sites. Obviously there is a wide range of potentials that could be considered, in order to produce more effective forces, and partial improvements will be expected.

The less is $p$, the weaker is the effect of the interaction between the particles. For fixed $(\beta, \kappa)$ the flow is feasible only for those values of $p$ being less than a certain threshold $(\beta, \kappa)$ dependent. Above this threshold, the density becomes too high and the transmission process is prevented. We say the system is saturated. The saturation mechanism begins with the appearance of saturated domains, whose size grows collapsing the whole lattice.

Our purpose is to quantify this threshold density, as well as to describe the flow properties along the parameter space $(\beta, \kappa, p)$.

Such a model can be considered as a first step towards an *abstract* modeling of packet/message telecommunication networks. In such networks, packets are routed by each node according to routing tables dynamically maintained by the network so as to minimize some *cost function* along the trajectory of the packets [8]. In the present model, this cost function is simply the traveled distance, and, as the network state does not change during simulations, routing tables need not be dynamically calculated. A natural extension of this model would be to allow variable link lengths between the nodes and to compute routing tables according to the path lengths. The role of the temperature parameter is to quantify how firmly the network will try to stick to its policy of minimizing the cost along a trajectory: as will be shown below, implementing the routing according to the cost minimization in a rigid way (*zero temperature*), although the best *naive* choice from an individual user point of view, can be detrimental to the global behavior of the network and thus to its collective utility (the total throughput offered by the network).

Congestion is an unavoidable phenomenon in uncontrolled packet networks, since packets introduced at one node have no guarantee of finding the necessary resources (available buffer space) in the transit nodes. If uncontrolled, congestion results in packet losses which, in the case of data transmission, are detected by upper layers protocols (TCP for instance in the case of IP networks) and corrected by retransmission requests: packet losses may thus trigger more packets to be introduced in the network therefore amplifying the congestion. Two approaches are possible to avoid this *vicious circle*: reducing the source emission rate when a congestion is detected (this is the TCP approach, and it relies on the sources to behave differently to the congestion indications) or trying to control the congestion inside the network by preventing packets to access overloaded areas (this approach does not rely on any source behavior: in the extreme congestion case, the source is simply denied any access to the network). This last option, congestion control inside the network, is presently a very active area of research in the networking community [9]. To implement such a control, the overload information should be transmitted by a node to its neighbors at least. This is precisely what is modeled by the repulsive charge term of the present model: access to a loaded node is discouraged, and access to an overloaded node is simply forbidden. It should be noted that, in the model as it stands now, a node is instantaneously aware of the load state of its neighbors. A more complex model should take into account the latency introduced by the node-to-node transit time, leading to a description of the congestion control by some kind of *retarded potential*. It will be particularly important to take this latency aspect into account for the modeling of high speed networks [10] , but this is left for a future work.

# 3 Numerical Simulation

We have done Monte Carlo (MC) simulations in order to study the parameter space. We present results obtained in a $L \times L$ lattice with $L = 32$ and periodic boundary conditions. The computations have been carried out on Workstations.

The starting configuration is obtained by generating at each site a particle with probability $p$, therefore, the initial density is around $p \times L^2$. The random number generator is based on the one described in [11]. The time step is identified with a MC iteration, that is the update of $L^2$ lattice sites as described in section 2.

The temporal evolution of the system exhibits a transient regime, characterized by the instability in the observables (see next section). After this, the system falls in an extremely long-living metastable state, where the flow properties do not change significantly with the temporal evolution. We say the system has reached the asymptotic regime.

The time the system spends in the transient regime depends on the parameter space point. In this regime, in the non-saturated region, the particle density is initially low and grows with the particle injection.

This time also grows near the parameter space points where an abrupt change in the properties is observed (e.g. near the threshold density), reaching in this case up to $2 \times 10^4$.

At each value of the parameters, we have performed typically $8 \times 10^4$ MC iterations. For $\beta$ and $\kappa \in (0, 0.4)$ the transient regime takes around 5000 iterations, while above 0.5 for both parameters, this time falls to $400 - 600$ iterations. We have also made the simulations starting from different configurations, allowing the system to evolve until $2 \times 10^5$ iterations. We have computed the errors by calculating the dispersion between the results obtained starting from different configurations.

## 4 Observables

A correct description of the system is obtained from the measure of relevant observables. From their temporal evolution we will be able to know when the asymptotic regime is reached, or even if this regime will be or not saturated. From the averaged value of these magnitudes we will obtain a quantitative description of the flow process.

For a given configuration, we define the occupation $M$ as:

$$M = (1/V) \sum_{i=1}^{V} \sigma(i) \ . \tag{2}$$

The statistical average over the configurations (labeled by $j$) is:

$$\langle M \rangle = \lim_{N \to \infty} (1/N) \sum_{j=1}^{N} M_j \ . \tag{3}$$

being $N$ the number of averaged configurations.

During the time interval $t_j - t_i \equiv \triangle t$ , $n_f$ particles will reach their endpoint. We define the Band Width ($B_W$) as the number of particles arriving to the endpoint per time unit.

$$B_W(\triangle t) = \frac{n_f}{\triangle t} \ . \tag{4}$$

The statistical average for $B_W$ is obtained from its mean value over a number of time intervals $N_T$:

$$\langle B_W \rangle = \lim_{N_T \to \infty} (1/N_T) \sum_{j=1}^{N_T} B_W(j) \ . \tag{5}$$

We can obtain a good measure of the system performance, from the mean time employed by the particles for reaching their destination, $T_M$. We compute $T_M$ in the time interval $\triangle t$, by adding the individual time spent by each particle (delay time), divided by $n_f$:

$$T_M(\triangle t) = \frac{\sum_{i=1}^{n_f} T_i}{n_f} \tag{6}$$

In the same way, we define the statistical average as over $N_T$ intervals as:

$$\langle T_M \rangle = \lim_{N_T \to \infty} (1/N_T) \sum_{j=1}^{N_T} T_M(j) \tag{7}$$

The occupation frequency of a certain occupation number, $\sigma(n)$, is defined as the number of times that the occupation $\sigma(n)$ appears at any lattice site. The study of this quantity gives an indication of the required buffers

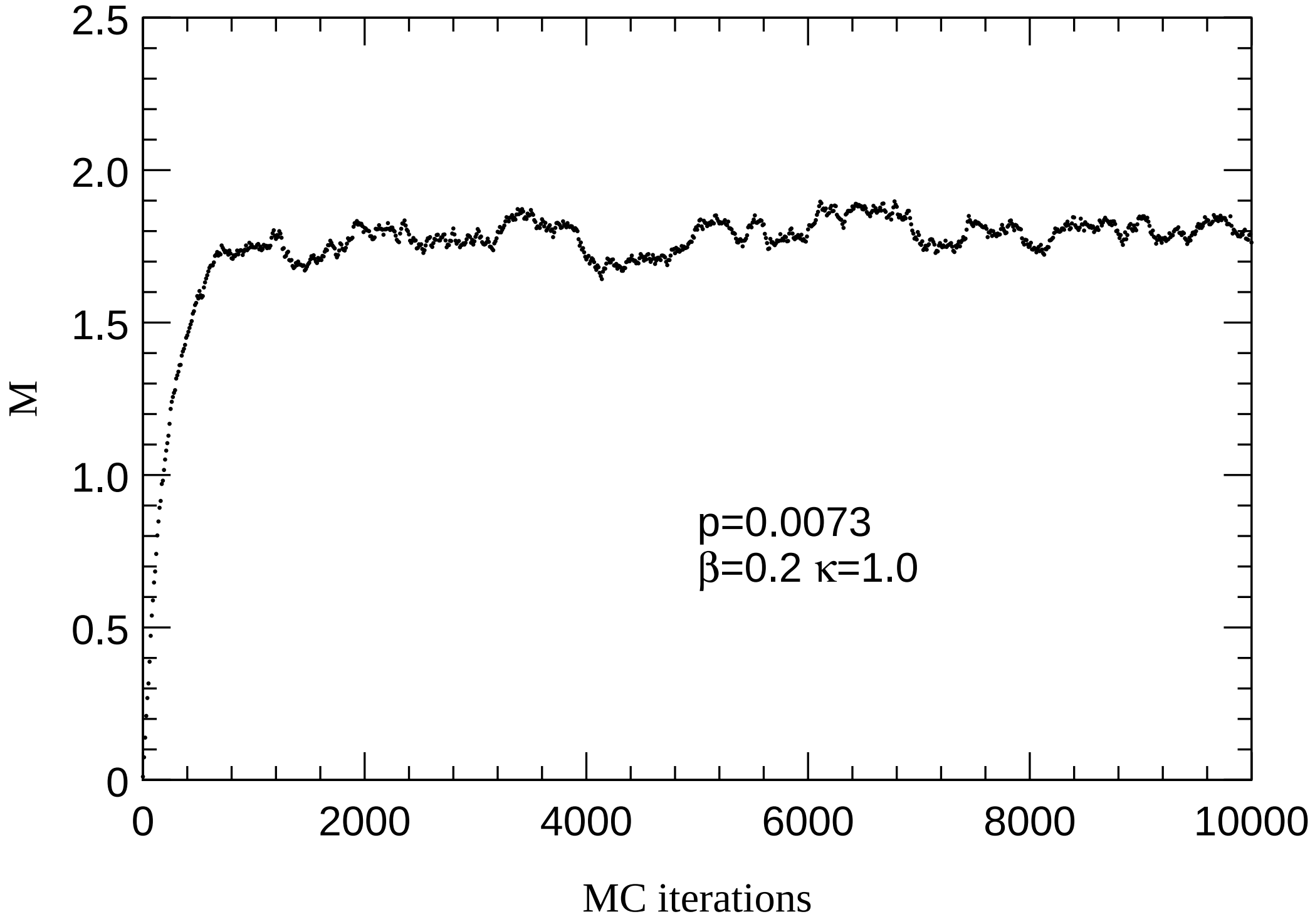

Figure 1: Temporal evolution of $M$ bellow the critical injection $p_c$.

# 5 Phase Diagram

We examine the parameter space $(\beta, \kappa, p)$ searching for regions where there are sharp changes in the temporal evolution. At each $(\beta, \kappa)$ value, there is a $p$ value denoted by $p_c$, such that for $p < p_c$ the asymptotic regime presents an stationary flow, and for $p > p_c$ the asymptotic regime is saturated and no flow is possible. We plot the temporal evolution of $M$ bellow $p_c$ (Figure 1) and above $p_c$ (Figure 2) for some values of the parameters.

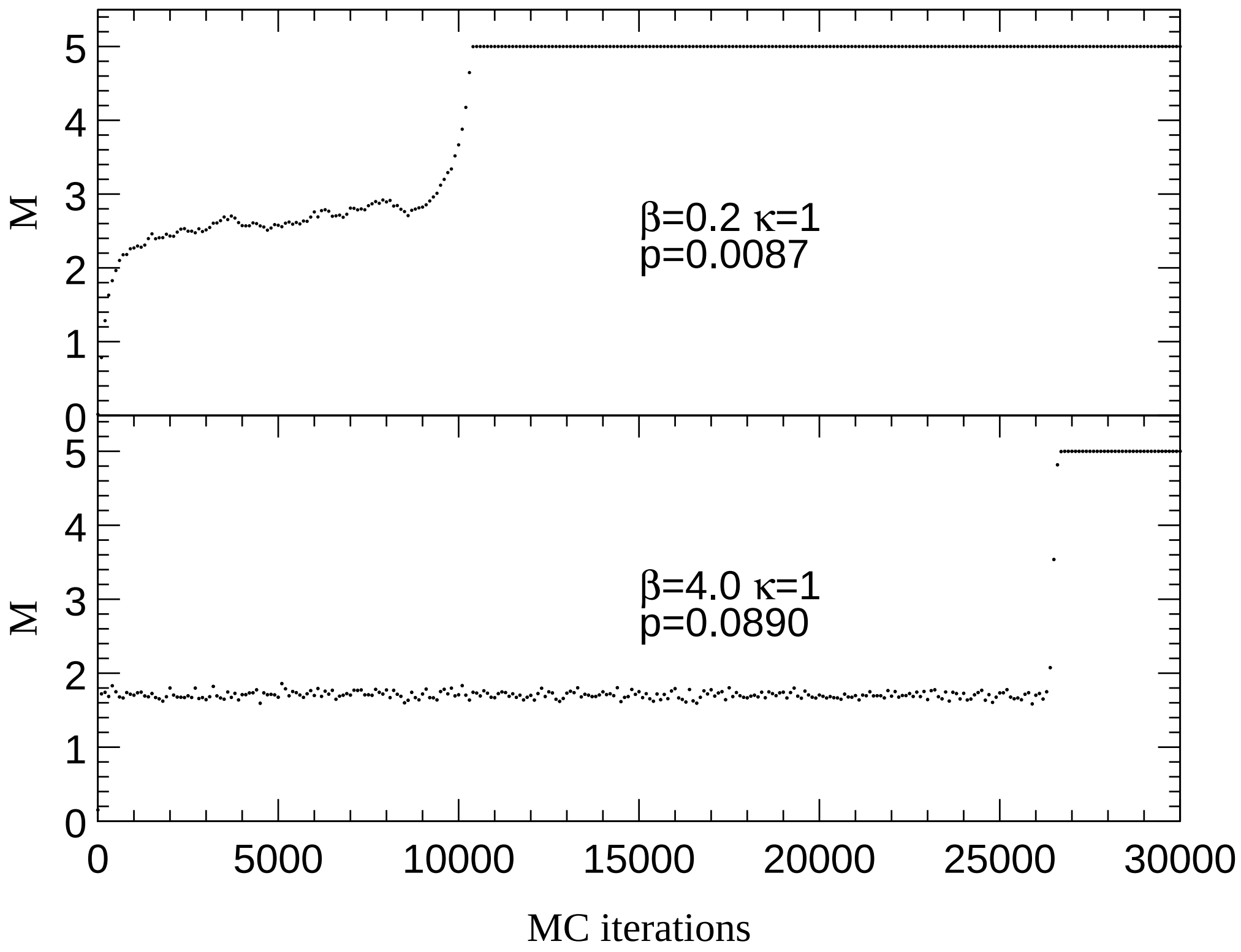

Figure 2: Temporal evolution of $M$ above $p_c$

This change is similar to a phase transition. The temporal evolution leads the system to one or another phase depending on the parameter space point. Once in the asymptotic regime, the non-saturated phase exhibits dynamic equilibrium: the number of injected particles, equals the number of arriving ones. This feature is reflected by $\langle B_W \rangle = p \times L^2$. In this phase, $\langle T_M \rangle$ is constant, as well as $\langle M \rangle$ which is always less than $B$.

The parameter space is divided in two regions by the surface defined by $(\beta, \kappa, p_c)$. Figure 3 shows two sections, for fixed $\beta$ and $\kappa$ respectively.

Above the surface, after the transient regime, $\langle M \rangle = B$, $\langle B_W \rangle = 0$ and $\langle T_M \rangle$ diverges.

It is possible to give a simple interpretation of this congestion phenomenon; when the load increases so does the probability that two particles residing on two neighboring saturated sites want to *exchange* their sites, but of course they cannot since one of the particles should move first and has nowhere to go as no resource is available at the desired site. This mutual blocking then tends to propagate to the other neighboring sites and above a given load, a complete congestion can develop at the scale of the network. Such a mutual blocking can also arise on a cycle of particles, although this is less probable for independent particles.

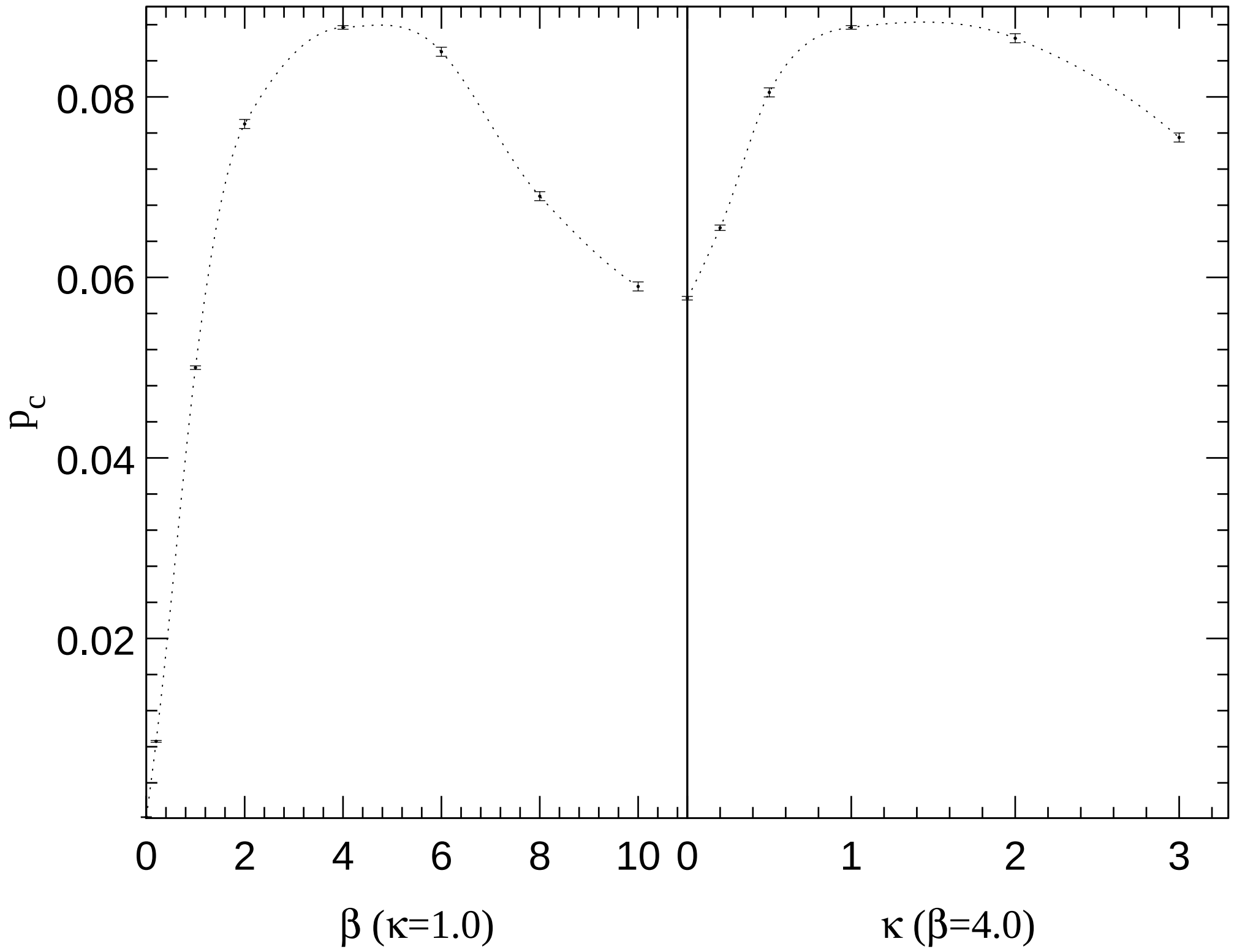

Figure 3: Sections through the phase space cube. On the left side is plotted $p_c$ versus $\beta$, on the right one, $p_c$ versus $\kappa$

Such mutual blocking is well-known in most *no loss* routing schemes such as *wormhole* routing and is called *deadlock* [12]. Deadlock avoidance is a very active research area, specially for massively parallel system interconnections [13].

### $\beta$ dependence

The thermal fluctuations move the particle away from its minimal distance path. The higher is $\beta$, the less important are these fluctuations.

As shown in Figure 3 (left side), there is an interval of $\beta$ values where the reduction of fluctuations has a positive influence on the throughput: $p_c$ rises, as well as $\langle B_W \rangle$, while $\langle T_M \rangle$ decreases, as is shown is Figure 4.

When going to great $\beta$ values, the situation does not persist. The absence of thermal fluctuations damage the throughput because the particles are not able of surrounding obstacles. As consequence $p_c$ and $\langle B_W \rangle$ decrease. $\langle T_M \rangle$ does not appreciably change from its minimal value, corresponding to the infinite directionality one.

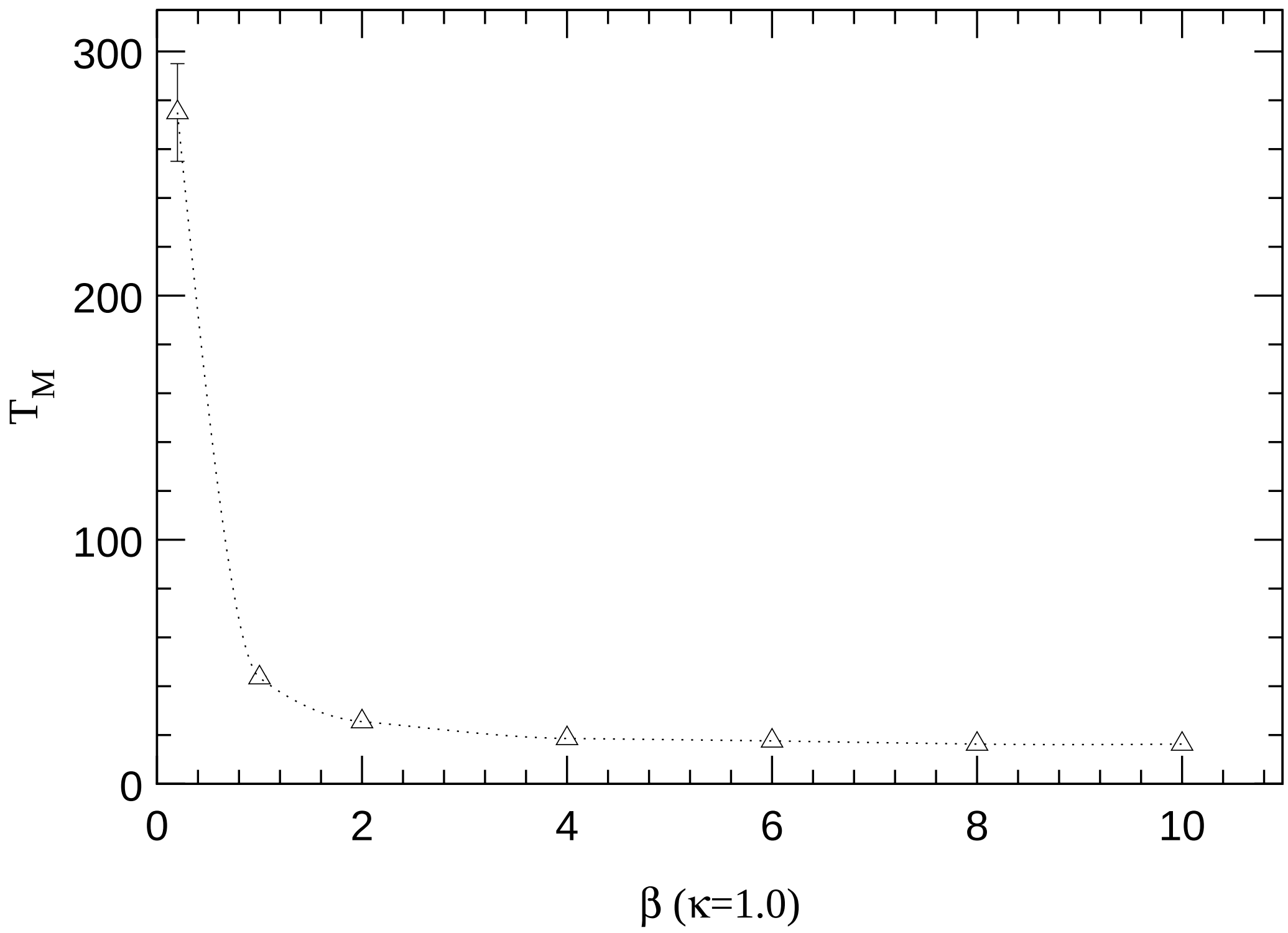

Figure 4: $\langle T_M \rangle$ dependence on $\beta$ for $p \approx p_c$

Figure 2 shows how the thermal fluctuations influence the saturation mechanism. If they are important, $M$ slowly grows until reaching $B$. If they are not significant, a sharp jump appears in the temporal evolution of $M$, between its value in the transient regime and $B$. We conclude from this that the saturated domains grow faster in absence of fluctuations, as could be expected from the earlier discussions.

In particular, it is intuitive that a mutual deadlock will last for a longer time if thermal fluctuations are disallowed (particles in mutual deadlock will repeatedly attempt to exchange their sites) hence creating a larger local congestion area which can eventually evolve towards a global congestion. This was clearly observed in [5] where only one buffer was available per site and this behavior also appears in this model, indicating that deadlock is more a consequence of a *no loss* routing scheme (movement is granted if and only if the available resource is available at the low end) than a consequence of insufficient resources.

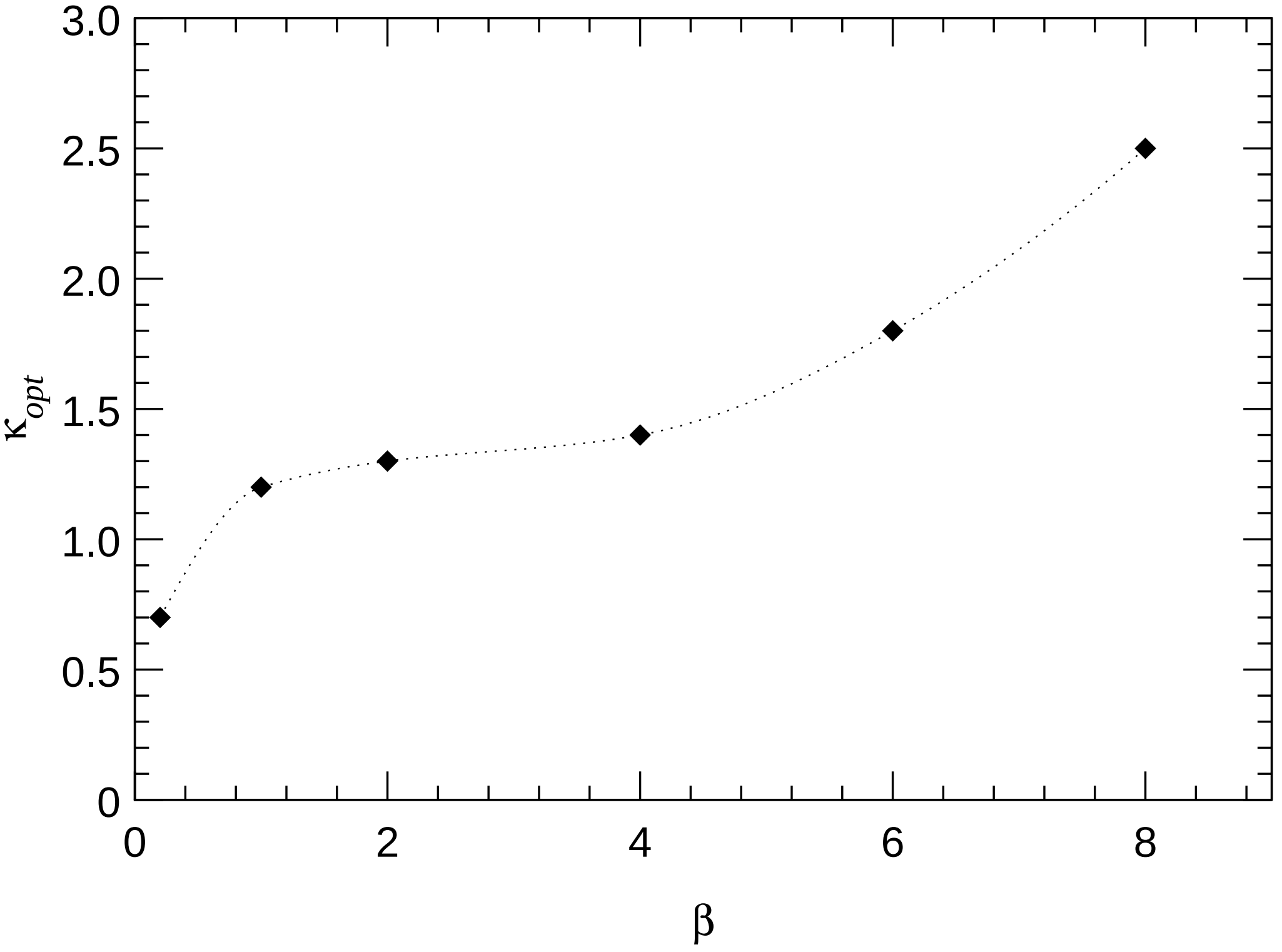

Figure 5: $(\beta, \kappa_{opt})$ line. The errors are of the size of the $\kappa$ step measured, $(\triangle\kappa = 0.5)$

**$\kappa$ dependence**

The dependence of $p_c$ on $\kappa$ presents two different regions (see figure 3).

In the first one, $p_c$ rises when increasing $\kappa$. For these $\kappa$ values, the inclusion of the repulsion term helps the system to avoid jammed regions. We denote by $\kappa_{opt}$ the value in which $p_c$ reaches the maximum value.

In the second region, $\kappa > \kappa_{opt}$, $p_c$ decreases with $\kappa$. From $\kappa_{opt}$ on, the repulsion is too strong and it moves the particles far away from their minimal paths. As a result, the collapse appears for smaller injection densities.

In Figure 5 we represent the value of $\kappa_{opt}$ for some $\beta$ values. The more restricted to its minimal path is the particle, the stronger is the repulsion needed in order to avoid obstacles in the lattice.

Therefore, there is a maximum value for the particle injection supported at each $\beta$ value, which is reached for $\kappa$ in the neighborhood of $\kappa_{opt}$.

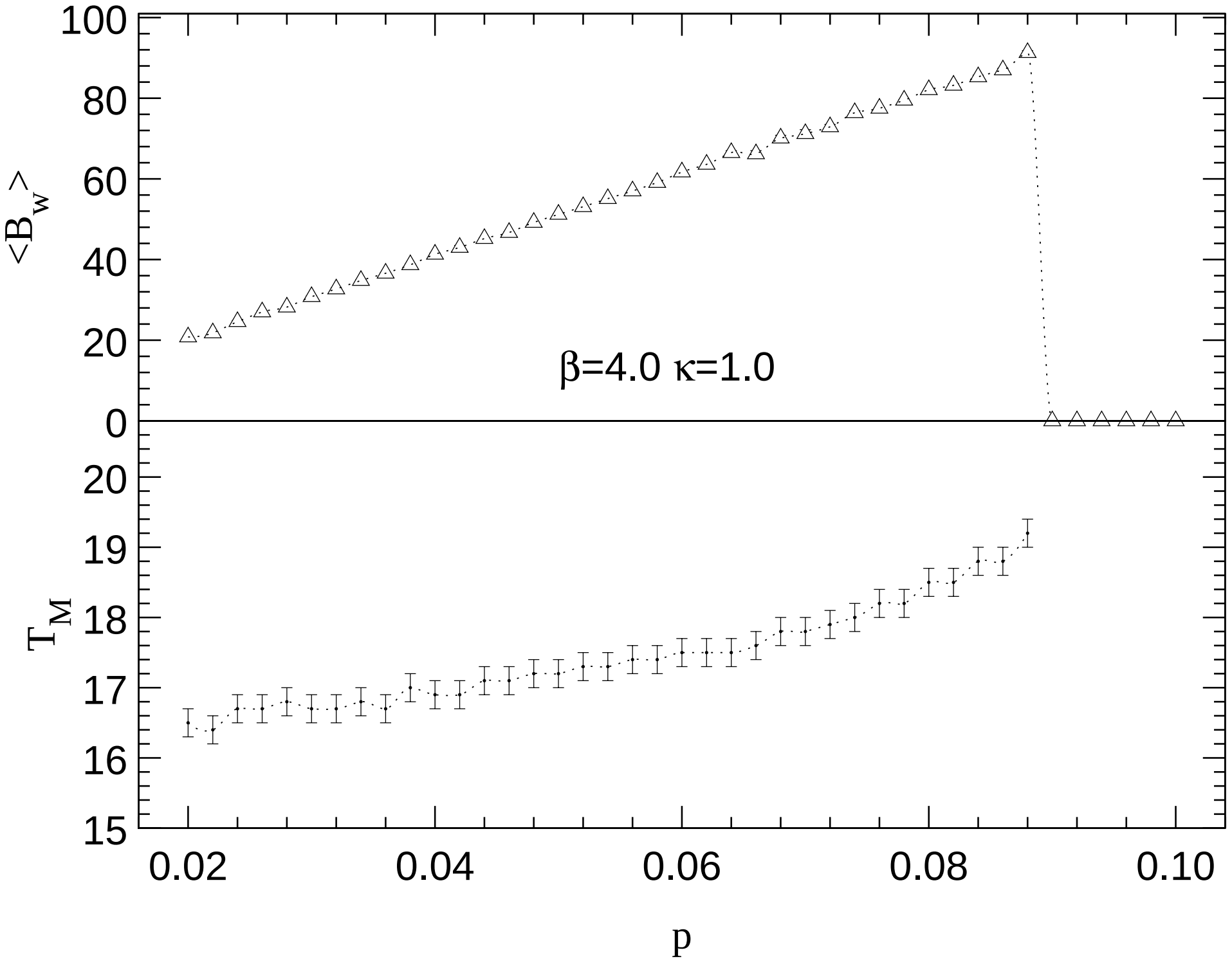

Figure 6: $\langle B_W \rangle$ (upper part) and $\langle T_M \rangle$ (lower part) versus $p$

# 6 General Behavior and Optimization

Let us focus on the behavior of the other significant observables for the traffic flow, when $\beta$ and $\kappa$ have been tuned in order to obtain the maximal $p_c$.

### Behavior of $B_W$

As we have already pointed out, $B_W$ in the asymptotic regime is proportional to $p$ by a factor $L^2$, as it may be checked in figure 6. So at each $(\beta, \kappa)$ value, the maximum for $\langle B_W \rangle$ is reached at $p = p_c$.

Above $p_c$, $\langle B_W \rangle$ remains constant during the transient regime, and sharply falls to zero after it. $B_W$ is a good measure in order to monitorize when the system has reached the asymptotic regime.

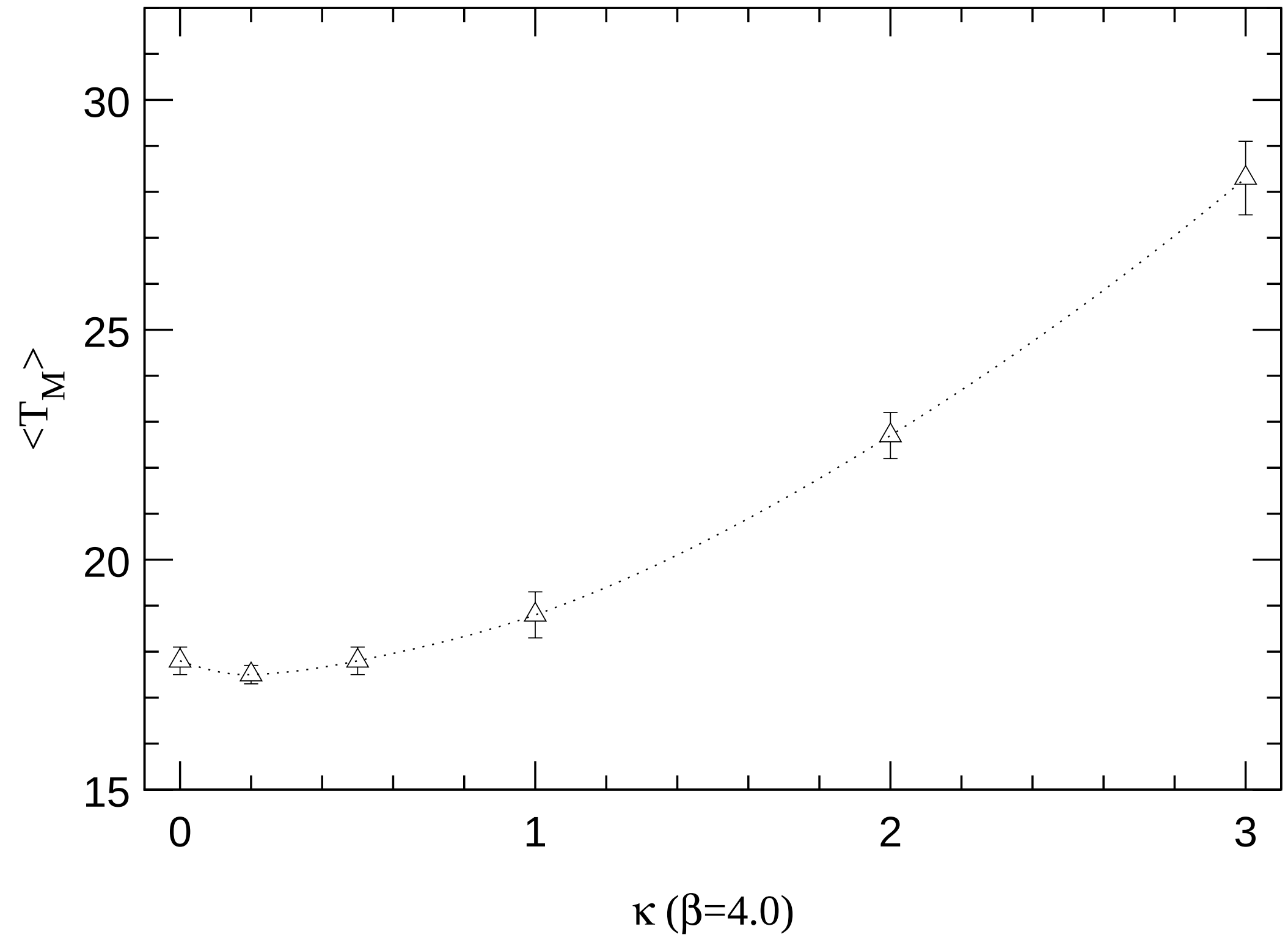

Figure 7: $\langle T_M \rangle$ dependence on $\kappa$.

**Behavior of $T_M$**

The greater is $p$, the stronger is the effect of the interaction over the particles, they more and more turn aside their minimal path increasing $\langle T_M \rangle$ (Figure 6). In Figure 7, we observe how $\langle T_M \rangle$ slightly increases with $\kappa$ until $\kappa_{opt}$. From $\kappa_{opt}$ on, $\langle T_M \rangle$ increases with a high slope.

We conclude that the better performances for the maximal injection supported, and for the bandwidth, are obtained on the line $(\beta, \kappa_{opt})$. The repulsion term damages $T_M$, the greater is $\kappa$, the more time spend the particles in reaching their destination. Anyway, it is only for $\kappa > \kappa_{opt}$ that $\langle T_M \rangle$ increases in a dramatic fashion.

Figure 8 shows the particle distribution in the simulation for $\beta = 0.2$ and $\beta = 4.0$. The plot of $\sigma(n)$freq$[\sigma(n)]$ exhibits a maximum around $\langle M \rangle$, and also reveals a wider distribution of particles for small $\beta$ values.

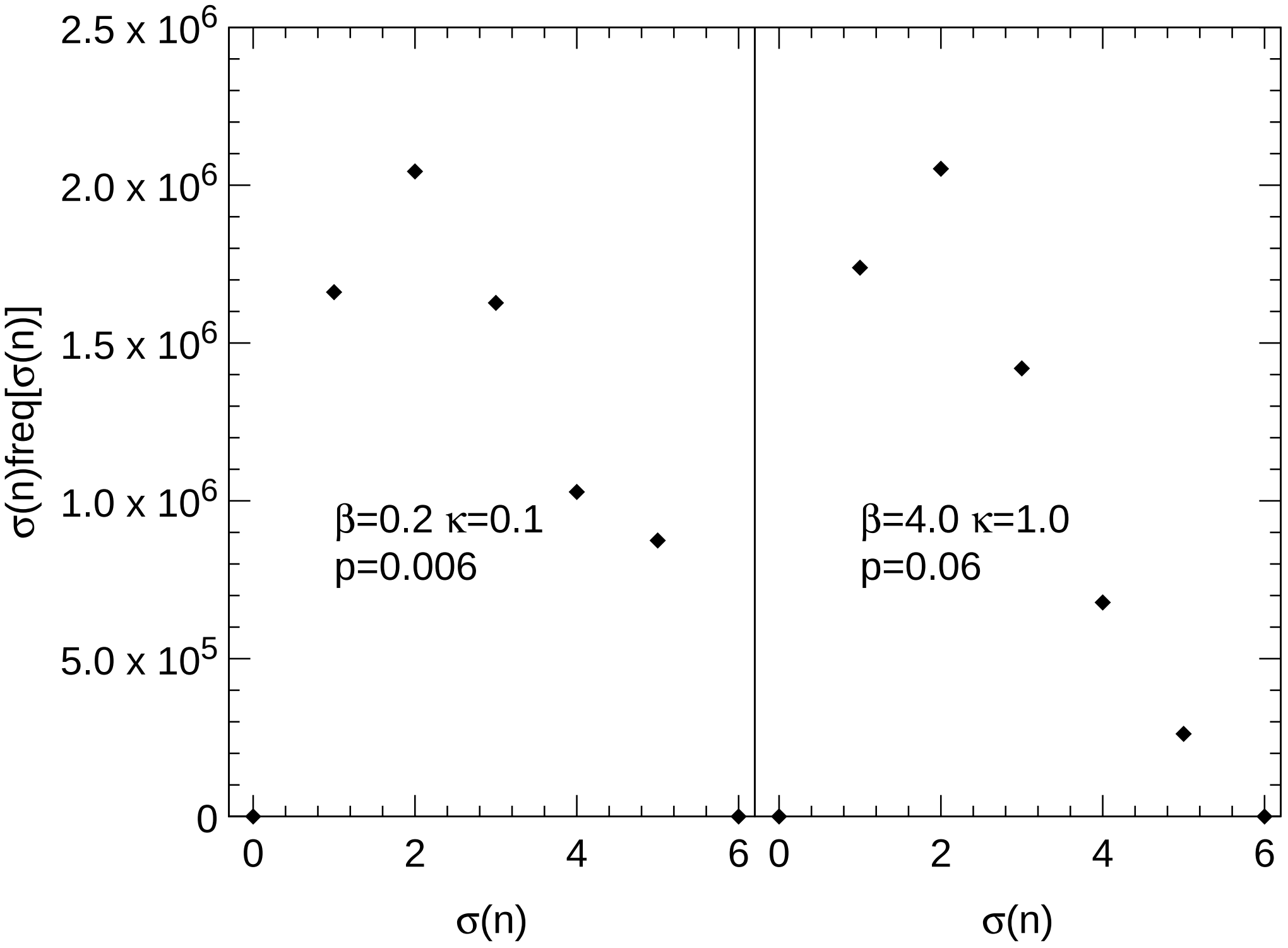

Figure 8: Distribution of the occupation number $\sigma(n)$ times the frequency of this state versus $\sigma(n)$ after $5 \times 10^5$ MC iterations. Transient regime contributions has been discarded.

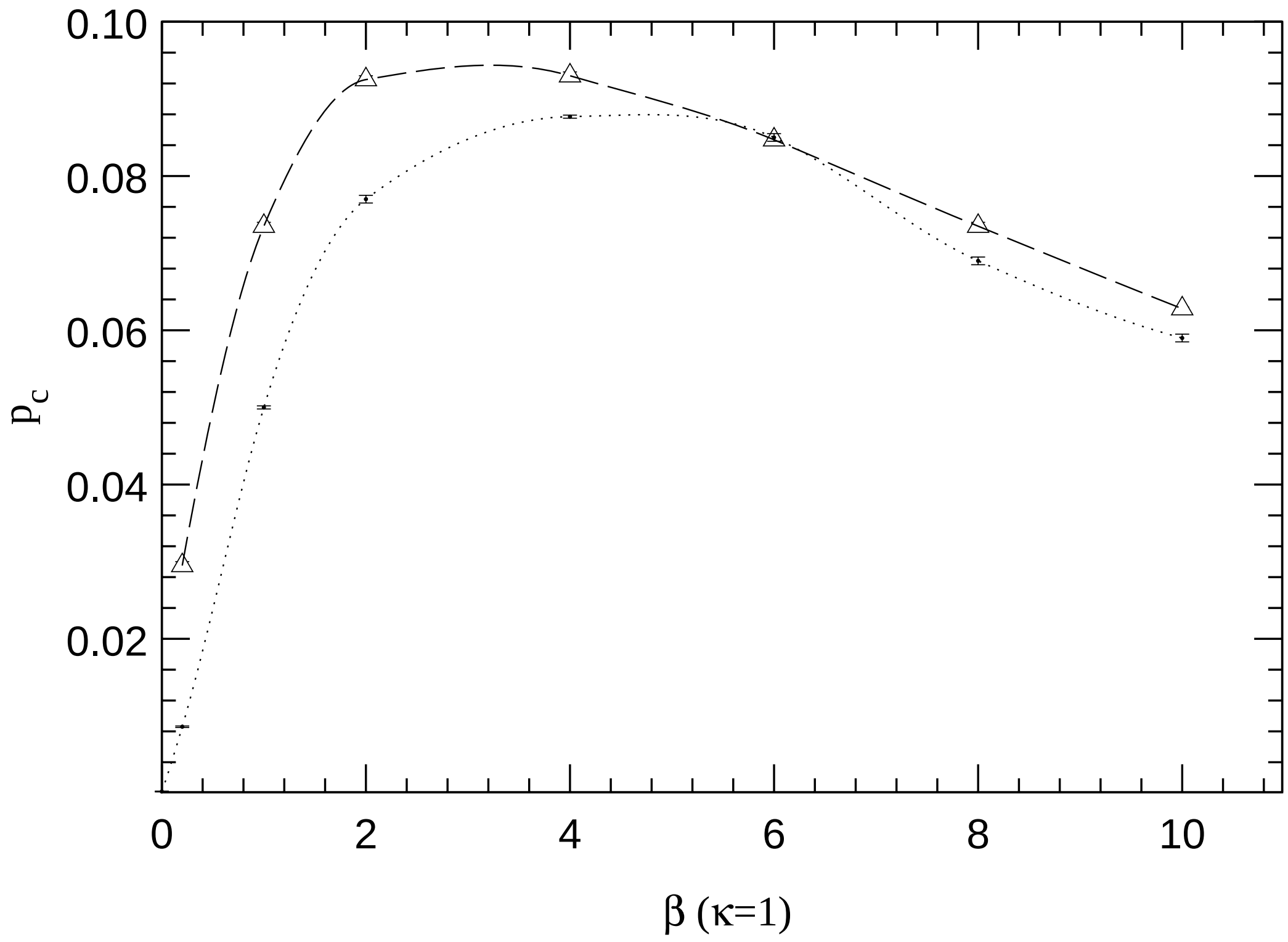

Figure 9: $\beta$ dependence of $p_c$ for the distance dependent force (dashed line). Dotted line represents the values obtained for the constant force

# 7 Improving the throughput: force depending on the distance

Although our implementation of the physical system has been inspired by the search for the simplest model accomplishing the desired features, certainly, by taking into account details not included up to now, partial improvements are to be expected. As an example, we develop a possible improvement. In particular, we deal with the quenching of thermal fluctuations when they are not useful anymore, that is, in the neighborhood of the endpoints.

We have seen how the $\beta$ term acts as a potential approaching the particles to their destination. In the primary implementation, we have used a constant force, and therefore the particles support thermal fluctuations of the same strength, no matter the remaining distance to the endpoint. We can expect a globally clearer lattice, if the particles distant a few sites from their destination are prevented of fluctuating. We will see how the inclusion of this feature do not spoil the good properties of the throughput, and also that the general behavior of the system is not altered.

A possible implementation to take this fact into account is obtained by the introduction of a distance dependence in the probability distribution, in such a way that the contributions of the fluctuations decrease when decreasing the distance to the endpoint.

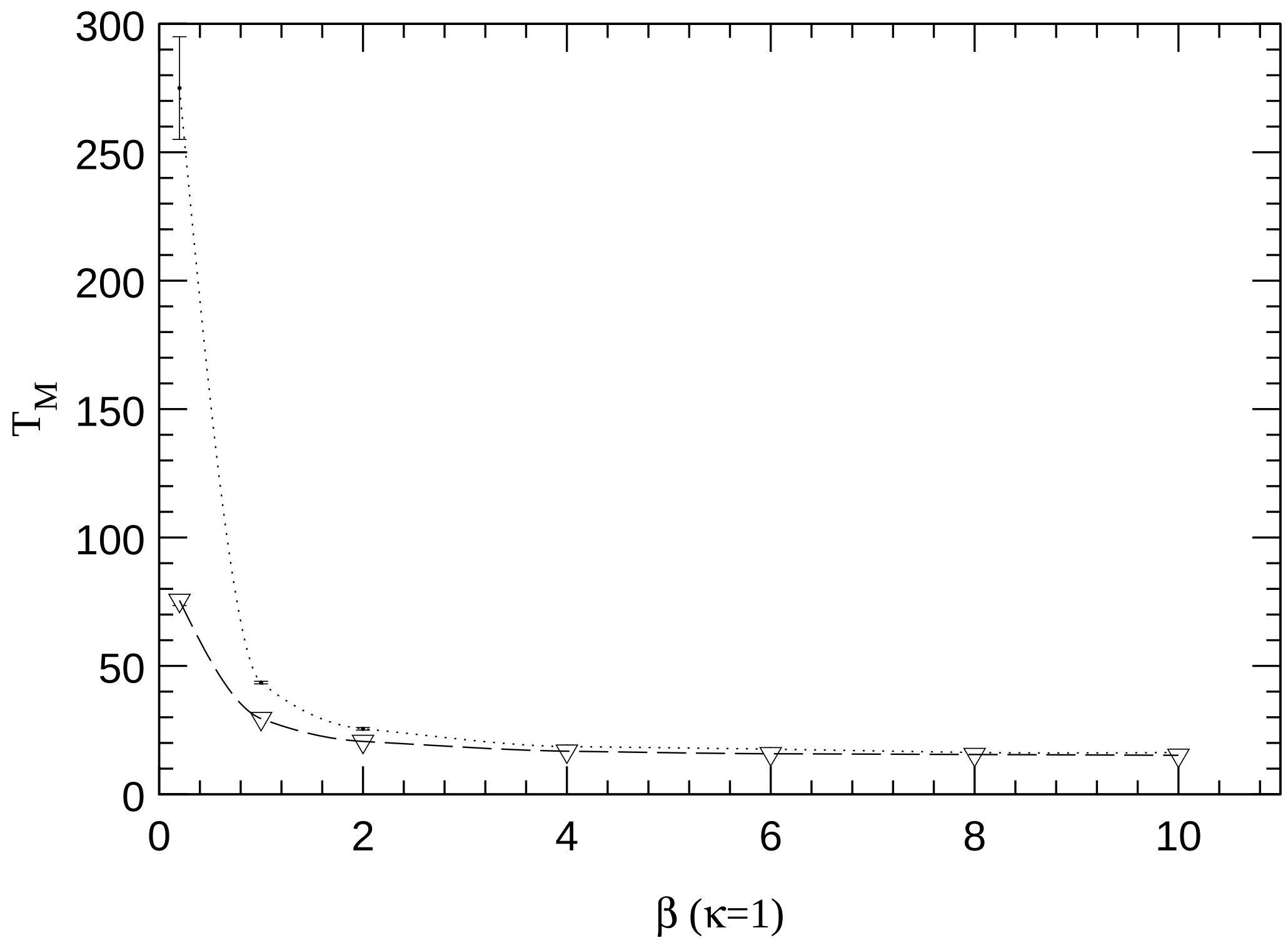

Figure 10: $< T_M >$ versus $\beta$ for fixed $\kappa = 1$ in $p \approx p_c$ for the distance dependent force (dashed line) and constant force (dotted line).

A straightforward way of doing that is to include a dependence on the relative approaching to the endpoint: The probability distribution reads now:

$$P(\pm\mu) = N \exp(\pm\beta \text{sign}({n^f}_\mu - n_\mu) - \kappa\sigma(n_\mu)) \left(\frac{r_n}{r_{n+\mu}}\right) \ . \qquad (8)$$

Where $r_n$ is defined by:

$$r_{(n_0,n_1)} = \sqrt{(n^f{}_0 - n_0)^2 + (n^f{}_1 - n_1)^2} \ , \tag{9}$$

In Figure 9 the evolution of $p_c$ with $\beta$ is shown. A global throughput improvement is reflected by higher $p_c$ values. This effect is more remarkable when the size of the thermal fluctuations is important (small $\beta$ values), corroborating our first intuition on the effect of fluctuations in the steps preceding the endpoint.

In Figure 10 we plot the $\beta$ dependence of $T_M$. We observe a global decrease in this time for all $\beta$ values.

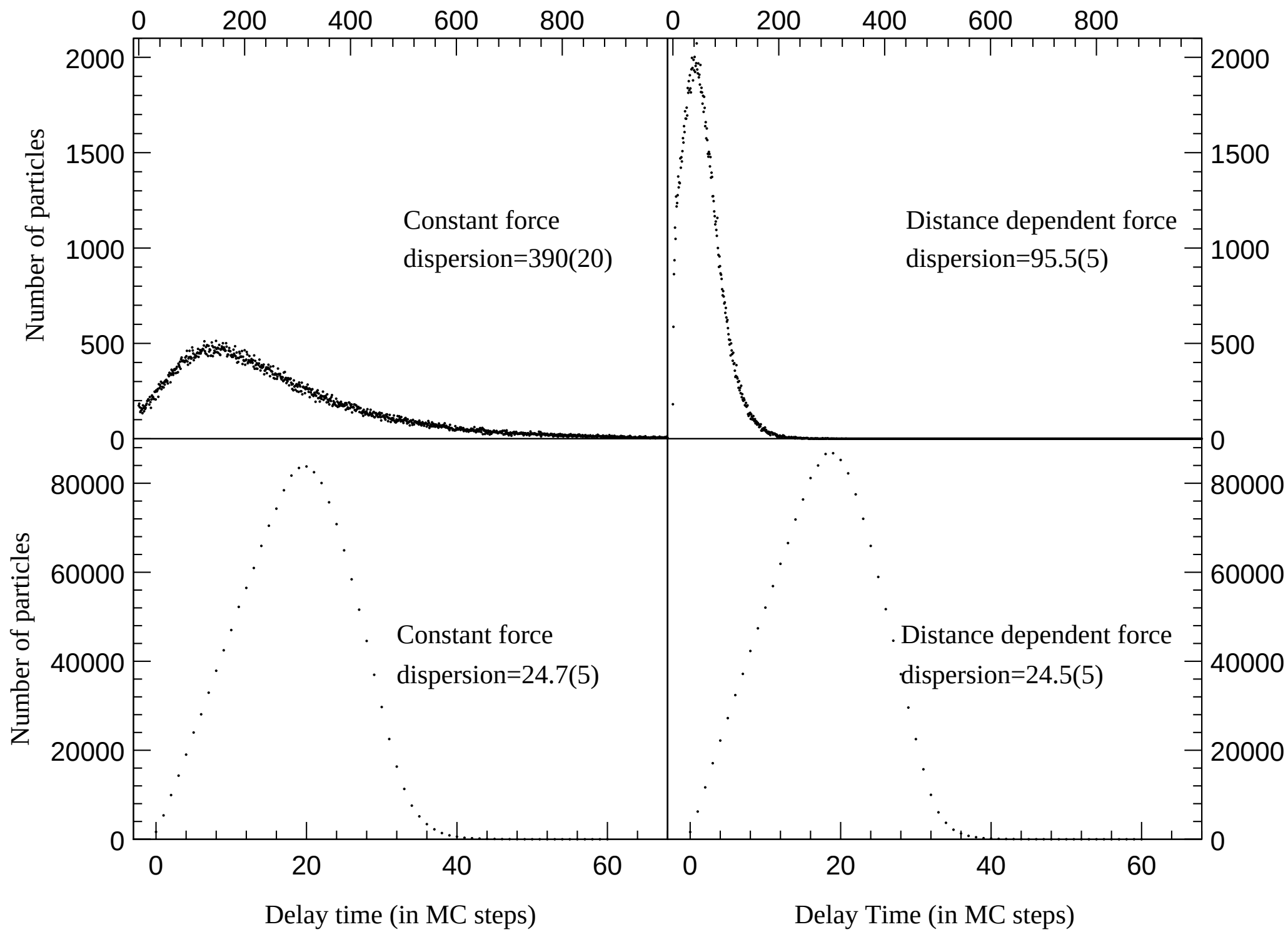

Figure 11: Delay distribution for the whole particles after $4 \times 10^4$ MC iterations. Graphics on the left (constant force) must be compared with the right side ones (distance dependent force). The distributions on the top are calculated in ($\beta = 0.2, \kappa = 1, p = 0.08$) on the bottom they correspond to ($\beta = 4, \kappa = 1, p = 0.008$).

The improvements concerning the delay time, are not limited to a smaller

$\langle T_M \rangle$. Figure 11 shows the delay distribution for all the particles, compared with the delay distribution obtained with the constant potential. We see at $\beta = 0.2$ how the dispersion of the distribution strongly decreases when using the distance dependent force. So, the particles arrive in more similar times, increasing the uniformity and the reliability of the traffic flow.

# 8 Fault tolerance

As we have pointed out, the lattice sites simulate the control nodes of an information flow system. Needless to say that the sites are an idealization of nodes in real systems, because the possibility of communication failures is not allowed. Real nodes are subject to external factors, as technical constraints or outside influences that often damage the communication ability between some nodes. These nodes are then temporarily out of order, and they cannot communicate information nor receive it from the other ones. It is then interesting to have an idea of how robust the system is when temporary problems in the transmission exist.

To implement communication problems, we prevent the information exchange with $n$ randomly chosen sites, during an interval $\triangle t \equiv 10\ MC$ iterations in the simulation. After this time interval, these $n$ sites are again allowed to communicate, and other $n$ are break at random [1].

In general the measured values for the observables will be shifted by an amount depending on $n$ and on the parameter region. For each $p$ value, we define the shift in the observable $O$ as is value, relative to the one obtained in the ideal system:

$$\triangle_r O(p) = \frac{O(p, n=0) - O(p, n)}{O(p, n=0)} \ . \tag{10}$$

We have study the influence on the relevant parameters of communication failures, for $n = 10$, 100 and 200 in $\beta = 0.2$ and $\beta = 4.0$.

In figure 12 is shown the comparative evolution of $\langle M \rangle$ for $n = 10$ and $n = 100$.

The first consequence of a bad transmission is that the system must support globally higher occupations. We measure smaller $\triangle_r p_c$ values the lower is $\beta$, because, as has been already shown, the lattice supports higher occupation numbers when the fluctuations are important.

[1] We work in this section with the probability distribution (1)

$\triangle_r\langle M\rangle$ is almost zero for both $\beta$ values, with $n = 10$ (around 1% of nodes out of order), while with $n = 100$, $200$ $\triangle_r\langle M\rangle$ is always large.

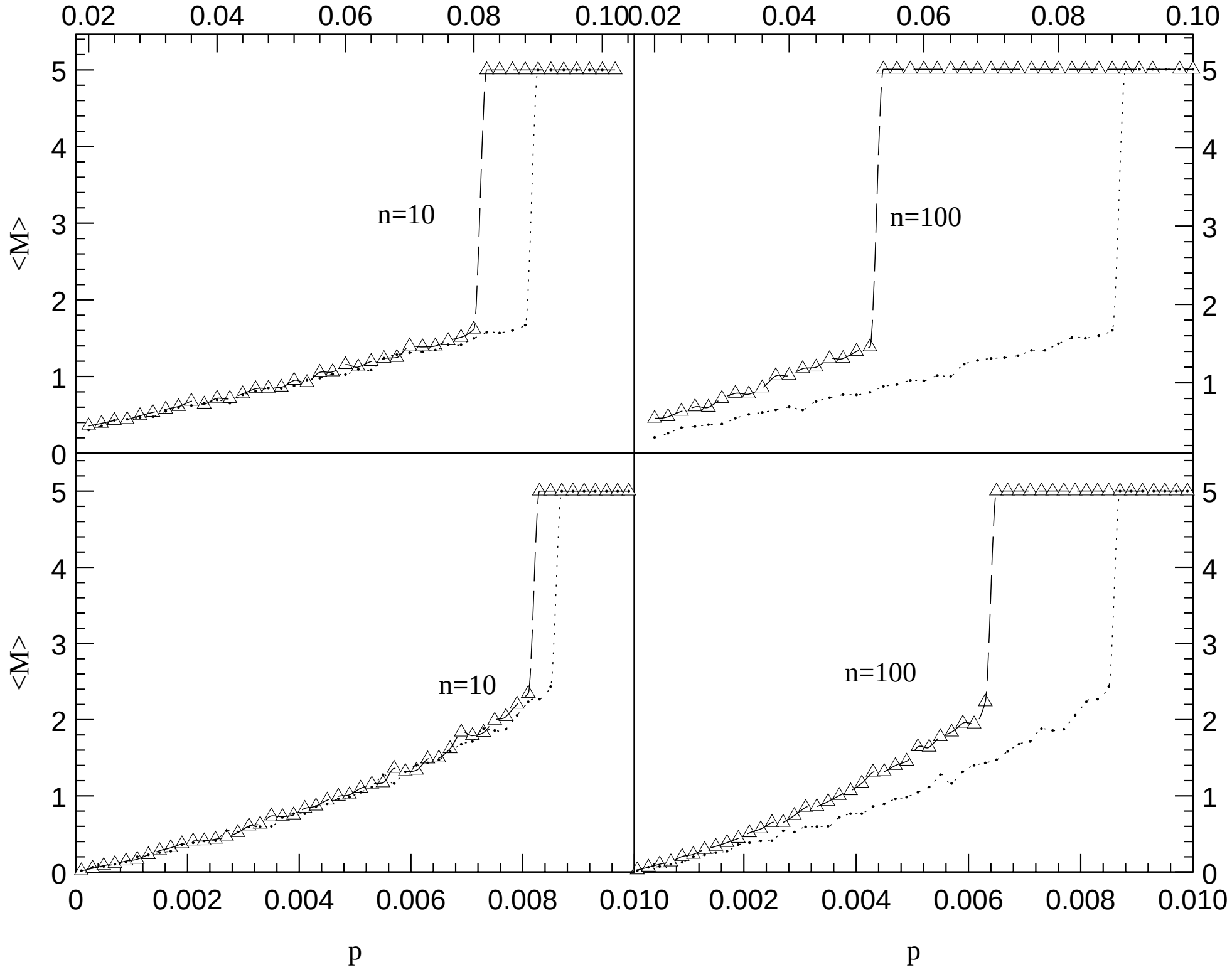

Figure 12: $\langle M\rangle$ versus p, for some values of the faulting percentage. Dotted lines represent the $\langle M\rangle$ values obtained for the ideal lattice. On the top the parameter space point is $(\beta = 4, \kappa = 1)$, on the bottom the plotted point is $(\beta = 0.2, \kappa = 1)$

In Table I we give the results obtained. As could be expected $\langle T_M\rangle$ increases for all $n$ values, even though, it is almost zero for $n = 10$.

# 9 Summary and Outlook

We have studied a model useful for describing the relevant processes occurring in flow information systems, with immediate applications to network message passing and traffic problems in general.

| Shift | $n = 10$ | $n = 100$ | $n = 200$ |
|---|---|---|---|
| $\beta = 4.0$ $\kappa = 1.0$ | | | |
| $\triangle_r p_c$ | 0.09 | 0.40 | 0.58 |
| $\triangle_r \langle T_M \rangle (\approx p_c)$ | -0.05 | -0.38 | -0.89 |
| $\triangle_r \langle M \rangle (\approx p_c)$ | 0.00 | 0.37 | 0.55 |
| $\beta = 0.2$ $\kappa = 1.0$ | | | |
| $\triangle_r p_c$ | 0.04 | 0.24 | 0.41 |
| $\triangle_r \langle T_M \rangle (\approx p_c)$ | -0.00 | -0.08 | -0.37 |
| $\triangle_r \langle M \rangle (\approx p_c)$ | 0.00 | 0.43 | 0.50 |

Table 1: Shifted values of the some relevant observables for some values of the number of failures $n$.

The introduction of the parameters $\beta$ and $\kappa$ as controllers of the system behavior allows us to go a step further from purely descriptive models, because we are able of giving prescriptions to improve the performance of the flow process.

Deeper studies, skipped in this first work are also possible. Concretely, an accurate study of the scaling with $L$ of the relevant magnitudes, as well as a detailed description of how the saturation time behaves, could lead us to the definition of quantities analogous to critical exponents. Also non-equilibrium states could be studied in order to monitorize the parameters controlling the saturation process.

The introduction of the $\kappa$ parameter has implied that the particles are able of avoid jammed regions. The study has been limited to contact interactions, because the particles only see the occupation of its nearer neighbors. By informing the particles about the occupation of wider surrounding regions, improvements in the throughput are to be expected.

**Acknowledgements**

This work has been partially supported by CICyT AEN93-0604-C03, AEN94-0218 and AEN93-0776. One of us (I.C.) thanks CERN grant, through the TSF program, and computer facilities, where part of this work has been done.

## References

[1] A. G. Greembergand, B. Hajek, *IEEE Transactions on Communication* **40**, 1070, (1992).

[2] J. A. Cuesta, F. C. Martínez, J. M. Molera and A. Sánchez, *Phys. Rev. E* **Vol. 48**, R4175 (1993).

[3] When-Shyan Sheu and Katja Lindenberg, *Phys. Lett. A* **147**, 437 (1991).

[4] O. F. Schilling, *Phys. Rev. B* **44**, 2784 (1991).

[5] I. Campos and A. Tarancón, *Phys. Rev. E* **50**,91 (1994).

[6] S. Mukherji and S. M. Bhattacharjee, *Phys. Rev. E* **48**, 3427 (1993).

[7] A. Sokal in *Quantum Fields on the Computer*, *World Scientific, Singapore*, (1992).

[8] D. Berteskas and R. Gallagher, *Data Networks*, *Prentice Hall, Englewood Cliffs*, (1987).

[9] F. Bonomi and K. W. Fendick, *IEEE Network* **9**, 25 (1995); H. T. Kung and R. Morris, *IEEE Network* **9**, 40 (1995); K. K. Ramakrishnan and P. Newman, *IEEE Network* **9**, 49, (1995).

[10] L. Kleinrock, *IEEE Communications Magazine* **30**, 36, (1992).

[11] G. Parisi and F. Rapuano, *Phys. Lett. B* **157**, 301, (1985).

[12] L. M. Li and P. K. McKinley, *Computer*, **February**, 62, (1993).

[13] D. J. Pritchard and D. A. Nicole, *IEEE Trans. Parallel and Distributed Systems* **4** 111 (1993).